# Demonstration of acoustic high-order Stiefel-Whitney semimetal in bilayer graphene sonic crystals

Xiao Xiang[1], Xiang Ni[2], Feng Gao[1], Xiaoxiao Wu[3,4], Zhaoxian Chen[5], Yu-Gui Peng[1,†], Xue-Feng Zhu[1,†]

[1]*School of Physics and Innovation Institute, Huazhong University of Science and Technology, Wuhan 430074, China*

[2]*Photonics Initiative, Advanced Science Research Center, City University of New York, New York, NY, 10031 USA*

[3]*Quantum Science and Technology Center and Advanced Materials Thrust, The Hong Kong University of Science and Technology (Guangzhou), Nansha, Guangzhou 511400, Guangdong, China*

[4]*Department of Physics, The Hong Kong University of Science and Technology, Clear Water Bay, Kowloon, Hong Kong, China*

[5]*College of Engineering and Applied Sciences, Collaborative Innovation Center of Advanced Microstructures, and National Laboratory of Solid State Microstructures, Nanjing University, Nanjing, 210023, China*

†Corresponding authors: ygpeng@hust.edu.cn (Y.-G.P.); xfzhu@hust.edu.cn (X.-F. Z.).

**Recently, higher-order topological phases have endowed materials many exotic topological phases. For three-dimensional (3D) higher-order topologies, it hosts topologically protected 1D hinge states or 0D corner states, which extend the bulk-boundary correspondence of 3D topological phases. Meanwhile, the enrichment of group symmetries with exploration of projective symmetry algebras redefined the fundamentals of nontrivial topological matter with artificial gauge fields, leading to the discovery of new topological phases in classical wave systems. In this Letter, we construct an acoustic topological semimetal characterized by both the first and the second Stiefel-Whitney (SW) topological charges by utilizing the projective symmetry. Different from conventional high-order topologies with multiple bulk-boundary correspondences protected by different class topological invariants, acoustic high-order Stiefel-Whitney semimetal (HOSWS) has two different bulk-edge correspondences protected by only one class (SW class) topological invariant. Two types of topological hinge and surface states are embedded in bulk bands at the same frequency, featuring similar characteristics of bound states in the continuum (BICs). In experiments, we demonstrate the existence of high-quality surface state and hinge state at the interested frequency window with polarized intensity field distributions.**

The everlasting goal of condensed matter physics is to search for new exotic matter, such as the topological insulators and semimetals [1-3]. The topological insulator is featured with the topological protection of transport of electrons on matter surfaces [4-5]. Encouraged by the existing exotic topological phases in 2D systems, such as the quantum Hall effect, quantum spin-Hall effect and Valley Hall effect [4-12], researchers began to explore new topological phases in higher-dimensional systems [13]. In the past years, the high-order topological semimetals have gradually gathered more attentions due to the existence of abundant topological phases [14-17]. In theory, for the high-order topology, a $d$-dimensional $n$th-order topological semimetal can support $(d−n)$ dimensional boundary states [18-20]. For example, 3D second-order and three-order topological topologies host the 1D hinge states and 0D corner states, respectively. With the unremitting efforts, 3D higher-order topological insulators and 3D topological semimetals have been demonstrated in various classical systems [16,17,21-25]. Up to date, these robust hinge states or surface states typically exist in the band gaps of topological insulators or discontinuities in topological semimetals, respectively, which were observed at different energy levels (or different frequencies) with different bulk-edge correspondences. Meanwhile, almost all high-order topologies with multiple bulk-boundary correspondences are protected by topological invariants of different classes [22-25]. A natural question arises as to whether there exists a topological phase that has more than one bulk-boundary correspondences but only one class topological invariant, with the different topological modes locating at the same frequency. Very recently, some important progresses were made for hunting this novel topological phase [26,27]. In the exploration of fundamental lattice groups, many extended algebraic relations can be projectively represented under the artificial gauge field [28-31]. Benefited from this approach, the 3D high-order Stiefel-Whitney semimetals (HOSWS) were proposed with four-fold or eight-fold degenerate nodal lines [27,32], in which the robust hinge states and surface states are protected by the same-class topological charges and thus coexist at the same energy level. However, the HOSWS are yet to be implemented in experiments and its inherent physical properties have not been fully investigated and failed to get a glimpse.

In this Letter, we report the realization of acoustic version of HOSWS by using the projective algebras with $Z_2$ artificial gauge field, which is a 3D second-order topological phase. The HOSWS hosts the 1D hinge states and 2D surface states that are embedded in the gapless band structure, which has similar properties of the bound states in the continuum (BICs) and indicated by nodal loops that possess two same-class topological charges (*viz.*, $w_1$ and $w_2$), where the topological charge $w_2$ is unique to the HOSWS. The creation of HOSWS is based on the extended $k \cdot p$ method [27], where we first generate the highly degenerate Dirac points (DPs) corresponding to high-dimensional irreducible representations. Next, we break the translation symmetry in *x-y* plane of the primitive cells so that the degenerate DPs split to form the nodal rings. The HOSWS is then realized by judiciously constructing the bilayer graphene sonic crystal, where the strengths of positive and negative couplings are adjusted to break the symmetry in *z* direction. For the experimental demonstration, we successfully fabricate the sample and observe the hinge states and surface states at the same frequency window characterized by the nontrivial Stiefel-Whitney topological charges. Furthermore, we demonstrate the polarization property with selective field distributions of the hinge and surface states. Our findings reveal that HOSWS is a novel topological phase in the spinless systems, which has significance in potential applications such as robust high-quality-resonance-based sensing, lasing and others.

*Construction of the HOSWS*

Here we start with a 2D tight-binding model by purposely configuring the positive and negative couplings under the $Z_2$ gauge field, so that each flux block has a $\pi$ flux, as depicted in the left panel of Fig. 1(a). When we break the translational symmetry $L_y$ by making it a dimer structure along $y$ direction as illustrated in right panel of Fig. 1(a), the quadruple degenerated DPs will break, as shown in Fig. 1(b). The Hamiltonian core of the 2D lattice can be expressed as

$$H_1 = g_{01} * G_{01} + g_{02} * G_{02} + g_{31} * G_{31} + g_{32} * G_{32} + g_{10} * G_{10} + g_{20} * G_{20} + g_{13} * G_{13} + g_{23} * G_{23} \quad (1)$$

with

$g_{01} = \mathrm{Re}(H_{12}+H_{43})/2$, $g_{02} = \mathrm{Im}(-H_{12}+H_{43})/2$, $g_{31} = \mathrm{Re}(H_{12}-H_{43})/2$,

$g_{32} = \mathrm{Im}(-H_{12}-H_{43})/2$, $g_{10} = \mathrm{Re}(H_{13}+H_{42})/2$, $g_{20} = \mathrm{Im}(-H_{13}+H_{42})/2$,

$g_{13} = \mathrm{Re}(H_{13}-H_{42})/2$, $g_{23} = \mathrm{Im}(-H_{13}-H_{42})/2$, $G_{01} = \sigma_0 \otimes \rho_1$, $G_{02} = \sigma_0 \otimes \rho_2$,

$G_{31} = \sigma_3 \otimes \rho_1$, $G_{32} = \sigma_3 \otimes \rho_2$, $G_{10} = \sigma_1 \otimes \rho_0$, $G_{13} = \sigma_1 \otimes \rho_3$, $G_{20} = \sigma_2 \otimes \rho_0$,

$G_{23} = \sigma_2 \otimes \rho_3$, $H_{12} = t + te^{-ik_x}$, $H_{13} = J_1 + J_2 e^{-ik_y}$, $H_{42} = -J_2 - J_1 e^{-ik_y}$, $H_{43} = t + te^{ik_x}$,

where $J$ and $t$ are the hopping strengths in $y$ direction and $x$ direction, respectively. $\sigma$ and $\rho$ denote the Pauli matrices acting on the row and column of the Hamiltonian matrix. Then we extend the 2D lattice into a bilayer graphite lattice with $\pi$ flux in each plaquette which is perpendicular to $z$ direction, as shown in Fig. 1(c). By breaking the parity symmetry along $z$ direction, the DPs at the six corners of the Brillouin zone split and form doubly degenerate nodal loops. The Hamiltonian core of the HOSWS lattice reads

$$H_2 = x_1 G_{01} + x_2 G_{02} + H_z, \quad (2)$$

where $x_1 = \text{Re}(ue^{-ia_1k} + ue^{-ia_2k} + ue^{-ia_3k})$, $x_2 = \text{Im}(ue^{-ia_1k} + ue^{-ia_2k} + ue^{-ia_3k})$, $Q_{13} = Q_1 + Q_2 e^{ik_z}$, $Q_{24} = -Q_2 - Q_1 e^{ik_z}$, $H_z = \begin{bmatrix} 0 & H_t \\ H_b & 0 \end{bmatrix}$, $H_t = \begin{bmatrix} 0 & Q_{13} \\ Q_{24} & 0 \end{bmatrix} = H_b^*$, where $Q$ and $u$ are the inter-layer and intra-layer amplitudes, respectively. By diagonalizing the Hamiltonian, the bilayer band structure can be calculated and displayed in Fig. 1(d). Each real DP splits into a real nodal loop, as denoted by the blue ring, which is characterized by two Stiefel-Whitney topological charges, as depicted by the yellow and red rings in the right panel of Fig. 1(d).

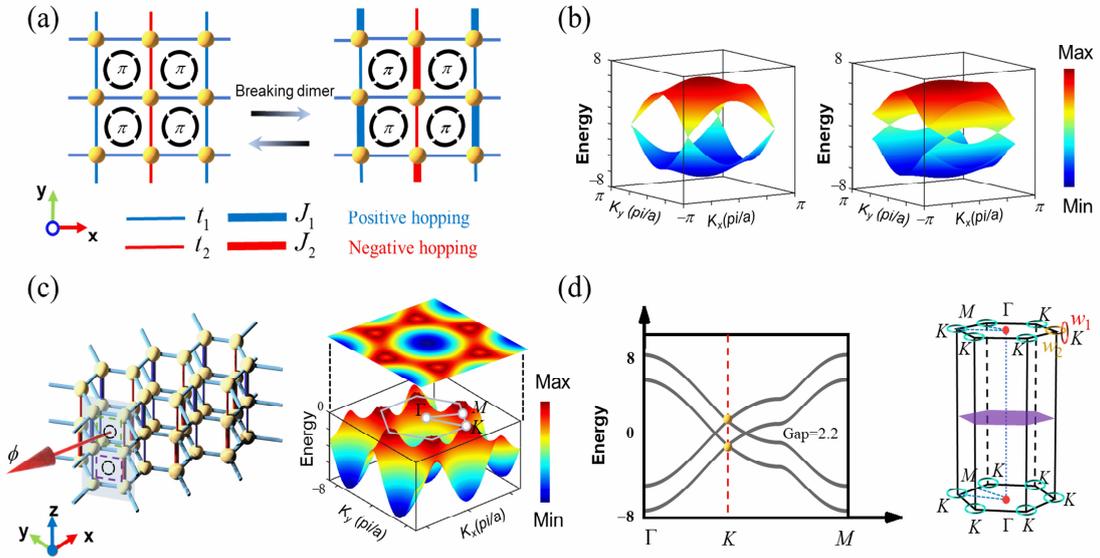

Fig. 1. The 3D higher-order semimetal with Stiefel-Whitney charges. (a) A 2D square lattice with $\pi$ flux in each unit cell. The blue and red channels connecting the adjacent atoms indicate the positive and negative couplings, respectively, where the width of channels indicates the strength of couplings. When $t_1 \neq t_2$, the lattice has a dimer structure along $y$ direction. (b) The band structures of the 2D square lattices for which $t=0.5$ and $J_1=1.3$ $J_2=0.7$. (c) The model of HOSWS, where each unit cell has four sites inside connected by intra-layer positive hoppings of strength $u$ and inter-layer hoppings of strengths $Q_1$ and $Q_2$ (the blue and red hoppings indicate positive and negative hoppings respectively), where $u=2.1$, $Q_1=1$, $Q_2=2.1$. The right panel shows the band structure of HOSWS, with degenerate nodal line loops on the zero energy plane. (d) The band structure of HOSWS along the high symmetric line of the first Brillouin zone. The right panel shows the first Brillouin zone where blue circles indicate the position

of nodal line rings, and the red and orange circles perpendicular to each other indicate the two Stiefel-Whitney charges of $w_1$ and $w_2$.

*Acoustic realization of the HOSWS*

For the acoustic realization of HOSWS, we employ the 3D bilayer graphene sonic crystal that comprises acoustic resonators and coupling tubes, which is sketched in Fig. 2(a). In principle, we can flexibly manipulate the signs and amplitudes of the coupling strengths, which are completely determined by the geometries of resonators and tubes. As shown in Fig. 2(a), $N$ bilayer graphene structures are stacked layer by layer in the $z$ direction, with the inset clearly illustrating the detail configuration of one unit cell. In the designed HOSWS acoustic model, the height of hexagonal cavities in the primitive cell $h$=30 mm, the inter-layer spacing $l_w$=16.5 mm, and the lattice constants along $x$ and $y$ directions $a_x = (\sqrt{3}L_w + 3R_0)/2$ mm, $a_y = (3\sqrt{3}R_0 + 3L_w)/2$ mm. For each cell, there exist four different sets of coupling tubes, for which the detailed geometric parameters can be found in Supplementary Materials. In order to generate the required positive and negative hopping pattern, we connect the intra-layer cavities with bending tubes, and connect the inter-layer cavities via vertical tubes with and without the space coiling, as illustrated in Fig. 2(b). To investigate the property of HOSWS, we calculate the frequencies of eigenstates of the sonic crystal via a finite element solver in Fig. 2(c). The purple dots represent the theoretically degenerate hinge and surface modes at the same frequency of 5510 Hz, where the red and blue dots represent the slightly split hinge and surface modes in a finite structure with the frequencies at 5515 Hz and 5506 Hz, respectively. The intensity fields of Stiefel-Whitney-charge protected hinge states and surface states in acoustic HOSWS are displayed in Fig. 2(d). In the simulation, the source is set at the center of the top layer. By conducting the Fourier transformation, we can further extract the pressure field on the top layer to calculate the dispersions of hinge and surface states, which are plotted by the thermal diagram in Fig. 2(e), in good agreement with the tight-binding calculation (the white-line band structure).

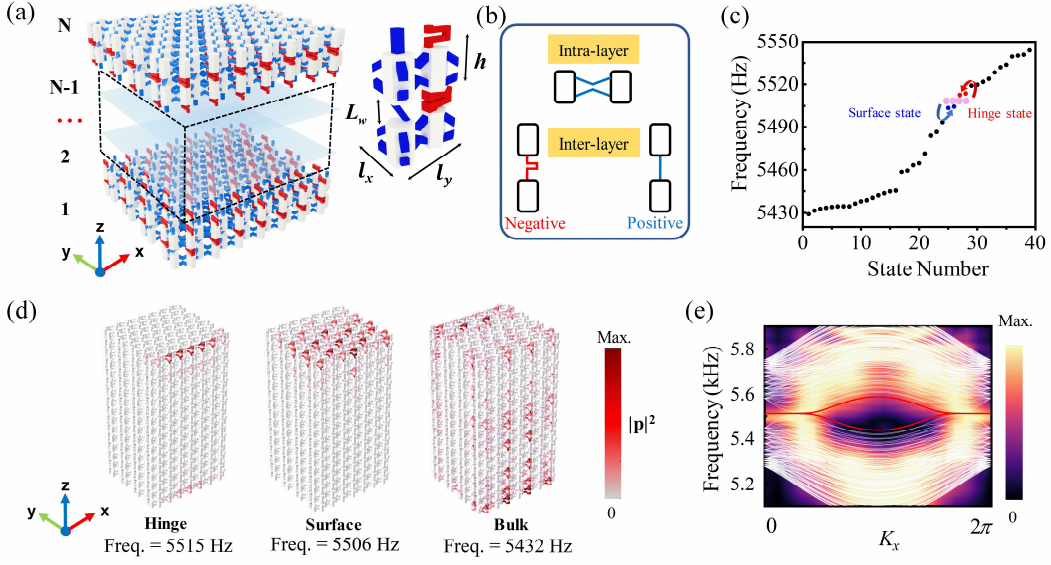

Fig. 2. Acoustic realization of HOSWS. (a) A bilayer graphene sonic crystal comprising hexagonal acoustic resonators and coupling tubes, in which the coupling tubes rendered by blue and red colors represent the positive and negative coupling, respectively. Inset: the enlarged configuration of one primitive cell. (b) The schematic of the negative and positive hopping between the intra-layer and inter-layer cavities. (c) The distribution of eigenstates of HOSWS calculated by a finite element solver. The purple dots represent the theoretically calculated hinge and surface modes, perfectly at the same frequency. The red and blue arrows denote the splitting of the modes in a finite acoustic structure, for which the red and blue dots represent the hinge and surface modes after splitting. (d) The intensity fields in HOSWS at 5515 Hz, 5506 Hz, and 5432 Hz, indicating the presence of hinge states, surface states, and bulk states. (e) Band structure of HOSWS along $k_x$ direction (the white lines). The thermal diagram displays the dispersion curves calculated by Fourier transforming the simulated pressure field of hinge state.

*Polarization of hinge and surface states in the HOSWS*

There exist two types of topological surface states in acoustic HOSWS, namely, Type-A surface state and Type-B surface state, as shown in Fig. 3(a). The intensity field distributions of Type-A and Type-B surface states reveal that the acoustic energies are selectively localized (or highly polarized) on the triangular-lattice sites and the inverted counterpart, respectively. Figure 3(b) show the rule of field localization for the surface states. The result reveals that the localization only depends on the sign of inter-layer coupling. To be specific, the sound energy only exist in the cavities which are linked by positive inter-layer coupling. Therefore, when we swap the positive and negative inter-layer coupling, the field distribution of surface state will change accordingly. Note that the field distribution of surface state is not homogenous on the top surface and is always inclined to be accumulated toward the boundary with positive inter-layer couplings. The polarization rule also holds for the topological hinge states, so that the hinge state only appear at the boundary with cavities connected by positive inter-layer couplings. As shown in Fig. 3(c), we show that, based on the polarization rule, we can manipulate the position of hinge state to the bottom, left, top, or right boundary respectively via the interlayer coupling engineering.

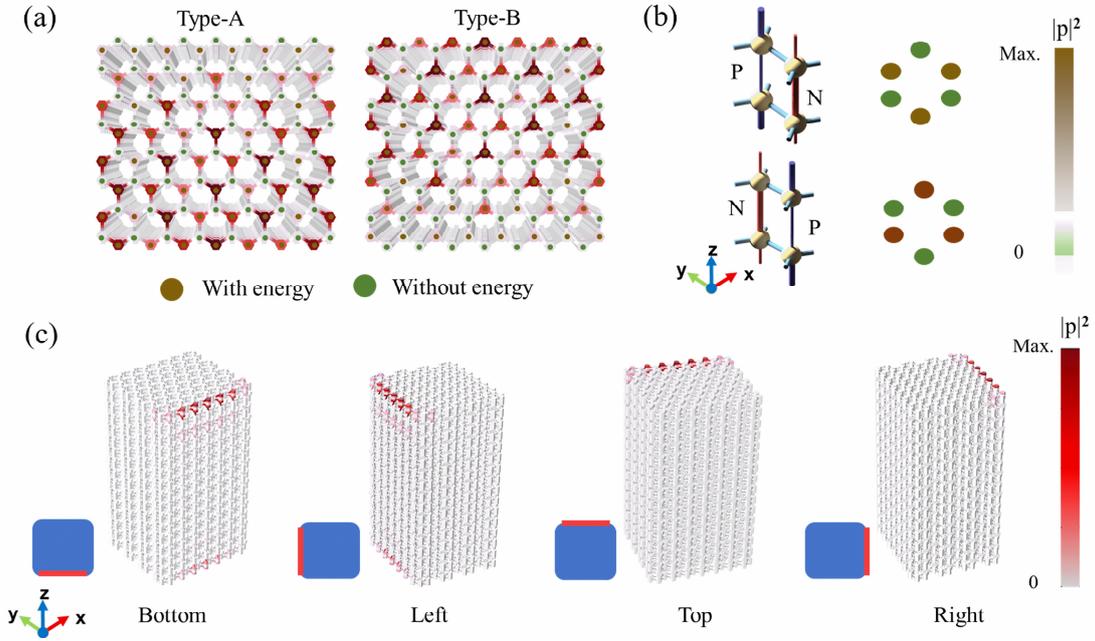

Fig. 3. Field polarization of hinge and surface states in HOSWS. (a) Type-A and Type-B surface states on the top layer of the lattice, where the acoustic energies for the two surface states are selectively localized at the yellow sites. (b) The intensity distributions of sound in a unit cell, in which the acoustic energies of the two surface state are always localized in the cavities connected by positive inter-layer coupling. (c) Manipulating the locations of hinge states by varying the distribution of positive and negative inter-layer couplings.

*Demonstration for the acoustic HOSWS*

We utilize the 3D printing technology (accuracy of 100 μm) to fabricate the sample for the demonstration of acoustic HOSWS. In Fig. 4(a), the 3D structure has 7×7×8 unit cells, including 1920 acoustic resonators. Two sources (Source 1 and Source 2), denoted by the red and blue stars on the top and bottom layers of the sample, are employed for the excitations of hinge and surface states. The right panel of Fig. 4(a) shows the schematic of one layer stacking block, where the measurement sites of Ports $P_1$, $P_2$ and $P_3$ are marked for characterizing the hinge, surface and bulk states. The experimental results for detecting the acoustic responses of the hinge, surface and bulk states are shown in Fig. 4(d). As expected, the spectrum for bulk states has an obvious dip at 5400~5600 Hz. However, for the measured spectra of hinge and surface states, there exist prominent peaks at 5520 Hz and 5503 Hz, respectively, in agreement with the result in Fig. 2(c). We can note that the signal peaks for hinge and surface states are at the same frequency window, where the slight position discrepancy roots in the finite size as well as the unavoidable sound absorption and sample deficit. To further confirm the property of topological hinge and surface states, we measure the intensity field distributions at the peak frequencies. As shown in Figs. 4(b) and 4(c), the intensity distributions on top and vertical surfaces at 5520 Hz and on top and bottom surfaces at 5503 Hz clearly present the features of hinge state and surface state, respectively. The measured field distributions verify the polarization effect, where the acoustic energy is selectively localized in the cavities connected with positive inter-layer couplings. More details for the experiments are appended in the Supplementary Materials.

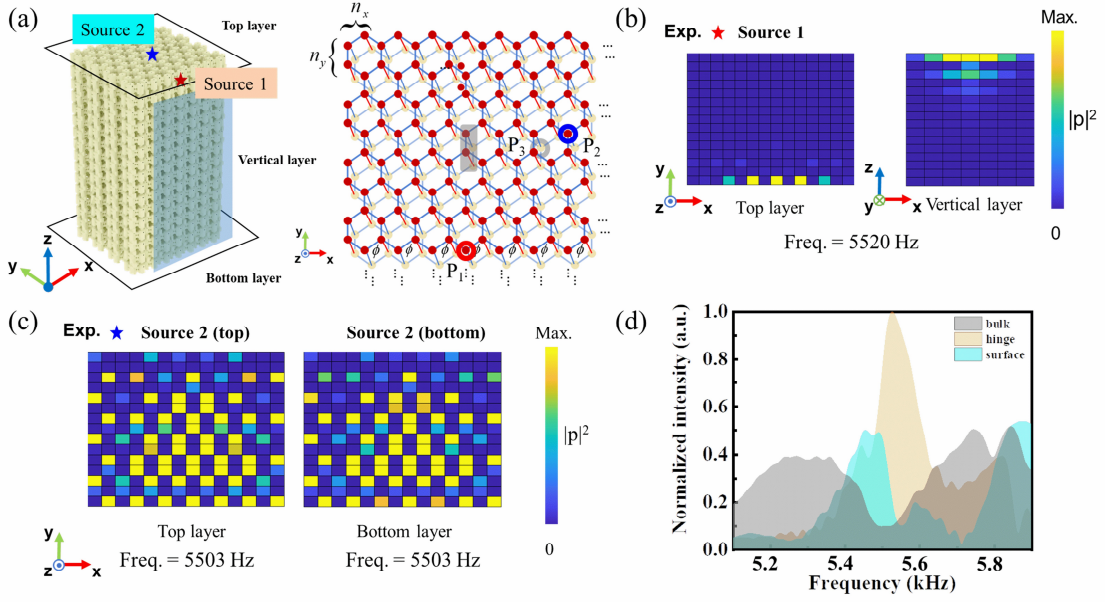

Fig. 4. Demonstration for the acoustic HOSWS. (a) The schematic of acoustic HOSWS. The top layer, bottom layer, and vertical layer are highlighted. In experiments, the sound source is set at Source 1 and Source 2 on the top/bottom layer, as marked by the red and blue star, aiming for exciting the hinge and surface states. Right panel shows the schematic of one single-layer building block with the measurement sites marked. (b) The transmission spectra measured at Ports $P_1$, $P_2$ and $P_3$ as marked by the red, blue, and gray circles in the right panel of (a), which characterizes the hinge, surface and bulk state. (c) The experimentally measured acoustic intensity field distributions on the top layer and vertical layer, for characterizing the hinge state excited by Source 1. (d) The measured acoustic intensity field distributions on the top layer and bottom layer, for characterizing the surface state excited by Source 2.

*Conclusion and outlook*

We demonstrate a novel topological phase of higher-order Stiefel-Whitney semimetal (HOSWS), for which two different bulk-boundary correspondences with only one-class topological invariant are realized in a time-invariant spinless system. The fundamental algebra of the symmetry group for the HOSWS is governed by the projective parity-time symmetry in the $Z_2$ artificial gauge field. By our judicious design, we utilize the bilayer graphene sonic crystal to construct the acoustic version of HOSWS. The topological phase is experimentally measured and visualized. The hinge and surface states are identified at the same frequency window. In addition, we discover that both the hinge state and surface state are polarized in the field distributions, for which the acoustic energies are selectively localized in the cavities connected with positive inter-layer couplings, allowing for the controllable topological localization via the inter-layer coupling engineering. Our work has significance in the exploration of new projective-symmetry-based topological phases, for which the different high-quality resonant states governed by the same-class topological invariant are important in the applications such as sensing, lasing and others.

*Acknowledgments*

The authors acknowledge financial support by the National Natural Science Foundation of China (Grant Nos. 11690030 and 11690032).

*Author contributions*

X. Xiang, F. Gao, X. Ni and Y. G. Peng developed the theory and did the simulations. X. Xiang designed the experiments and fabricated the samples. X. Xiang and F. Gao performed the experiments. X. X. Wu gave important advices on the simulations and experiments. All authors analyzed the data and wrote the manuscript. X. F. Zhu and Y. G. Peng supervised the project.